\documentclass[%
 reprint,https://www.overleaf.com/project/61d57ef6a28cc20966a60ed7
 amsmath,amssymb,
 aps,
]{revtex4-2}

\usepackage{graphicx}
\usepackage{dcolumn}
\usepackage{bm}

\usepackage{graphicx}
\usepackage{wrapfig}
 \usepackage{framed}
 \usepackage{color}
 \usepackage{float}
 \usepackage{epstopdf}

\makeatletter

\newcommand*{\rom}[1]{\expandafter\@slowromancap\romannumeral #1@}
\makeatother

\begin{document}

\preprint{APS/123-QED}

\title{Deforming Active Droplets in Viscoelastic Media}

\author{Prateek Dwivedi, Atishay Shrivastava, Dipin Pillai, Naveen Tiwari, and Rahul Mangal}
 \email{mangalr@iitk.ac.in}
\affiliation{%
 Department of Chemical Engineering, Indian Institute of Technology Kanpur, Uttar Pradesh-208016, India }%





\begin{abstract}
To mimic the motion of biological swimmers in bodily fluids, a novel experimental system of micellar solubilization driven active droplets in a visco-elastic polymeric solution is presented. The visco-elastic nature of the medium, characterized by the Deborah number ($De$), is tuned by varying the surfactant (fuel) and polymer concentration in the ambient medium. At moderate $De$, the droplet exhibits a steady deformed shape, markedly different from the spherical shape observed in Newtonian media. A theoretical analysis based on the normal stress balance at the interface is shown to accurately predict the droplet shape. With a further increase in $De$, time-periodic deformations accompanied by oscillatory transitions in swimming modes are observed. The study unveils the rich complexity in the motion of active droplets in viscoelastic fluids, which has been hitherto unexplored.    
\end{abstract}

\maketitle



Artificial active swimmers are capable of extracting energy from their surrounding and executing spontaneous mechanical motion ~\cite{RevModPhys.88.045006}. This non-equilbrium transport of soft matter systems has garnered significant interest due to their motion characteristics, which are markedly different from their Brownian counter-parts, and their potential to be used in a multitude of applications, including cargo-delivery in microscopic domains~\cite{doi:10.1021/nl072275j}, health-care ~\cite{GHOSH2020100836}, environmental remediation~\cite {doi:10.1021/nn301175b} and more. Unlike external field driven systems, these artificial swimmers are propelled by a self-generated local gradient of a physicochemical field such as temperature, concentration, surface-tension, etc.~\cite{doi:10.1080/03602548408068407} Active droplets are a class of active swimmers that self-propel due to the Marangoni stresses induced by interfacial tension gradient along the droplet interface ~\cite{C4SM00550C}. Spontaneous asymmetry is generated in an otherwise isotropic droplet with uniform surfactant coverage by either a targeted chemical reaction or by micellar solubilization.~\cite{C4SM00550C,doi:10.1146/annurev-conmatphys-031115-011517,Thutupalli_2011,DWIVEDI2022101614,doi:10.1021/la3015817} So far, experimental studies have been restricted to investigating the role of different parameters on the active droplets' motion, including droplet size~\cite {PhysRevLett.117.048003, PhysRevE.97.062703}, surfactant concentration gradient~\cite {doi:10.1073/pnas.1619783114}, external flow~\cite{doi:10.1063/5.0038716} and addition of external solute~\cite{doi:10.1063/5.0060952, PhysRevX.11.011043, PhysRevX.10.021035} in a simple Newtonian environment. A few theoretical and numerical studies ~\cite{doi:10.1063/1.4718446, PhysRevE.92.053008} have investigated the efficiency of different squirmer models in viscoelastic fluids. Very recently, Dwivedi \emph{et.al.} ~\cite{Dwivedi2022arxiv}, experimentally reported on puller-pusher-quadrupole mode transitions in  thermotropic liquid crystal 4-pentyl-4'-cyanobiphenyl (5CB) active droplets in ionic surfactant tetradecyltrimethylammonium bromide (TTAB) aqueous solution doped with Polyethylene oxide (PEO) as polymer. The study was however limited to weakly viscoelastic systems with low Deborah number, $De$, defined as the ratio of polymer relaxation time to the characteristic strain rate, i.e., $De=U\tau/a$. Here, $U$ is the droplet speed, $\tau$ is the polymer relaxation time and $a$ is the droplet radius.

In this letter, we investigate the active motion of droplets in ambient media with significant viscoelasticity, i.e, higher $De$. The self-propulsion of the oil droplets is achieved via micellar solubilization in an aqueous solution containing ionic surfactant. Unlike the so far explored Newtonian/weakly viscoealstic mediums, here we observe that the polymer solution produces a significant deformation of the motile droplets. Such large-scale deformation is attributed to the normal elastic stresses generated by the stretching of the polymer chains in the neighbourhood of the droplet interface.\\

 \vspace{-2mm}
\begin{figure}[H]
\centering
\includegraphics[width=0.9\linewidth]{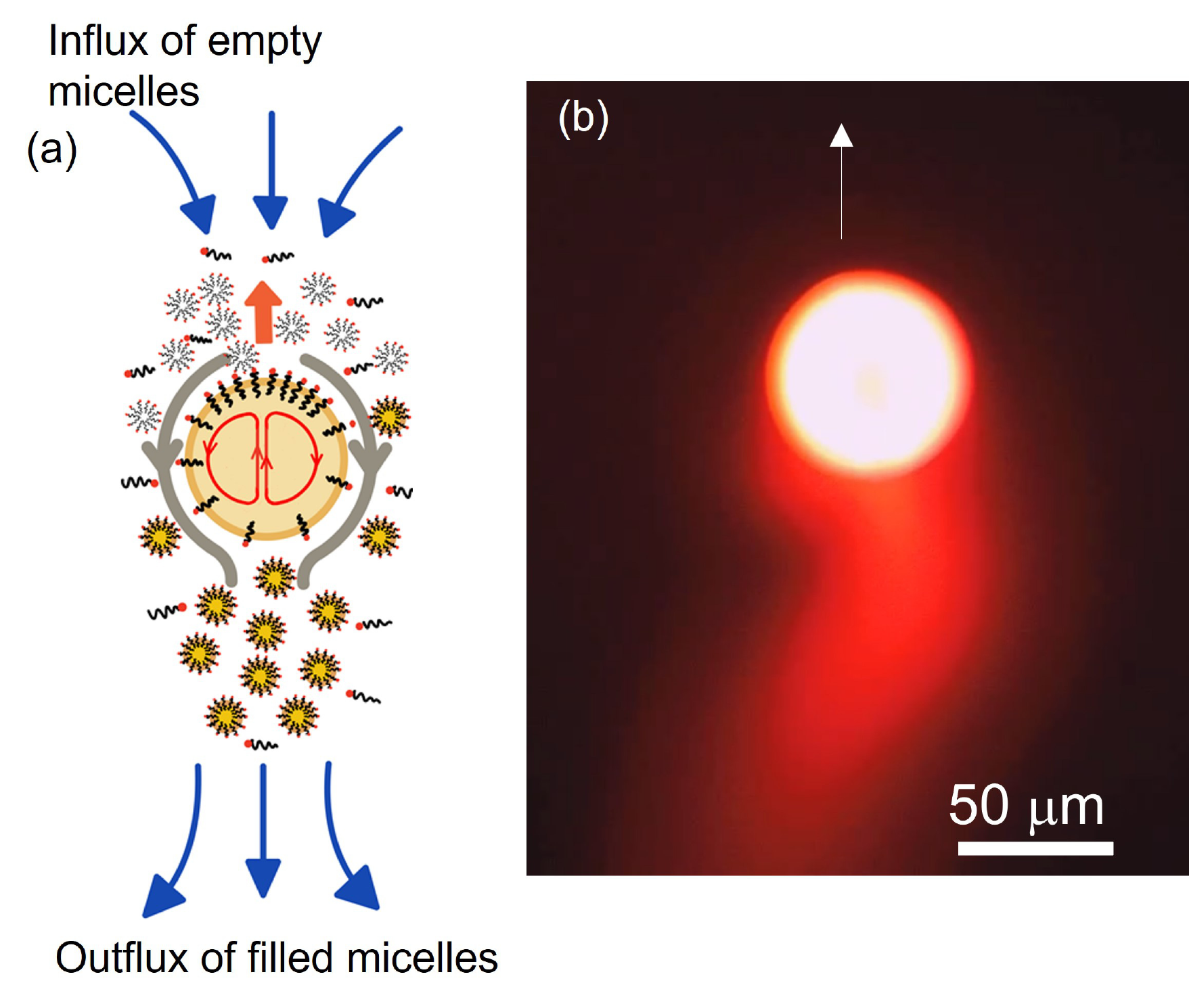}
\caption{(a) Schematic of micellar solubilization induced self-propulsion of droplets. (b) Fluoroscence micro-graph of an active 5CB droplet ($\sim$50 $\mu$m) tagged with Nile red fluorescent dye releasing a trail of filled micelles.}
\label{moving droplet}
\end{figure}

The experimental system consists of 5CB droplets ($\sim$50 $\mu$m)  dispersed in an aqueous solution of TTAB with various concentrations. In all cases studied, the surfactant concentration was kept far above the critical micellar concentration (CMC) (for TTAB in water 0.13 wt$\%$) such that the droplets self-propel via micellar solubilization (see fig.\,\ref{moving droplet}). PEO of molecular weights either M{$_{w,1}$}= 1x10$^6$ Da or M{$_{w,2}$} =8x10$^6$ Da were added to the solution, in varying amounts, to make different viscoelastic solutions. Prepared solutions were subjected to small amplitude oscillatory shear experiments to obtain the associated zero-shear viscosity $\eta_{o}$ and the variation in $G$' and $G$" with respect to the imposed frequency($\omega$). The cross-over of $G$'($\omega$) and $G$"($\omega$) yields the relaxation (via reptation mechanism ~\cite{alma991023793349705251}) time, $\tau$, associated with the long polymer chains. Table \ref{table:1} describes the physical and chemical properties of the different surrounding media used in this study.
  \begin{table}[h]
    \caption{Description of physico-chemical properties of different surrounding media.}
\centering
 \begin{tabular}{c c c c c}
    \hline
    \vspace{0mm}
 Fluid & $c_{TTAB}$ & $M_{w,PEO}$ & $c_{PEO}$ & $\eta_o$ \\ [0.1ex] 
  & $wt\%$ & $Da$ & $wt\%$ & $Pa~s$\\ 
 \hline
\rom{1} & 6 & NA & NA & 0.001 \\
\rom{2} & 6 & 8x10$^6$ & 1 & 11\\
\rom{3} & 21 & 8x10$^6$ & 1 & 11 \\
\rom{4} & 21 & 1x10$^6$ & 4.2 & 13\\
\rom{5} & 21 & 8x10$^6$ & 1.25 & 28\\ [1ex] 
 \hline
\end{tabular}
\label{table:1}
\end{table}

A customized Hele-Shaw cell with the dimensions 1 cm x 1 cm and height of $\sim$100 $\mu$m was prepared using clean glass-slides. Initially, the cell was filled with the surfactant solution (with or without PEO) and then 5CB droplets were injected using an Eppendorf micro-injector. The injection volume was adjusted to achieve droplets of size $\sim$50 $\mu$m. Since the thickness of the optical chamber is only slightly larger than the droplet size, the droplets remain confined in the 2D X-Y plane during their active motion. In order to avoid droplet–droplet interaction, 5CB droplets were injected into the cell with a low number density. An upright polarized optical microscope, Olympus BX53, was used to observe the active motion of the isolated droplets. Bright field visualization of droplets was carried out using Olympus LC-30 camera with 1024 x 768 pixel$^2$ resolution at 20 fps and particle image velocimetry (PIV) experiments were performed using ORX-10G-71S7C-C (FLIR) camera with 3208 x 2200 pixel$^2$ resolution. The Hele-Shaw cell was placed on a thermal stage, attached with microscope, to maintain constant temperature (25{$^\circ$} C).

\begin{figure}[H]
\centering
\includegraphics[width=1\linewidth]{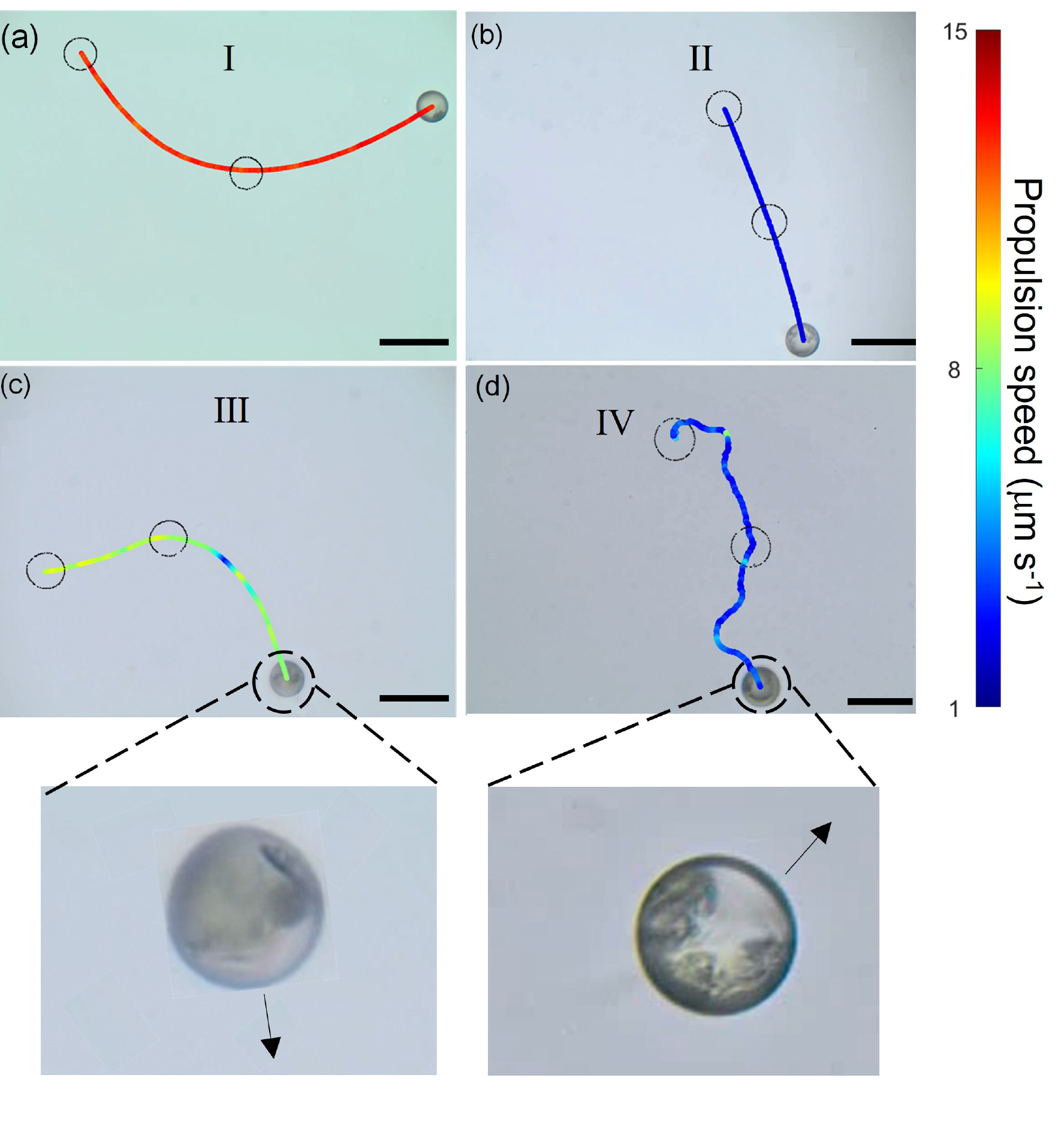}
\caption{Representative trajectories, color coded by the propulsion speed, of the active 5CB droplets in (a) Fluid \rom{1} (b) Fluid \rom{2} (c) Fluid \rom{3} (d) Fluid \rom{4}. The enlarged insets are the optical micrographs demonstrating the shape deformed shape of the droplet in (c) and undeformed shape in (d).}
\label{trajectories}
\end{figure}

First, we describe the characteristics of active motion of 5CB droplets in aqueous surfactant solution with c$_{TTAB}$=6wt$\%$ with no added PEO (Fluid \rom{1}). Consistent with the earlier reports on active motion in Newtonian media ~\cite {PhysRevE.97.062703}, at short time-scales the droplets expectedly exhibit ballistic motion (speed $\sim$ 15 $\mu$m s$^{-1}$) in different directions, maintaining their spherical shape. Next, we perform experiments with with 1 wt$\%$ PEO of M{$_{w,1}$}= 8 x 10$^6$ Da added to the surfactant solution, i.e., Fluid \rom{2}. Due to the increased viscosity ($\eta_o$=11 Pa~s) of the solution, droplets were observed to move slower (speed $\sim$ 2-3 $\mu$m s$^{-1}$). For this ambient medium, upon increasing c$_{TTAB}$ to 21 wt$\%$ (Fluid \rom{3}), the propulsion speeds were enhanced ($\sim$ 8-10 $\mu$m s$^{-1}$) and interestingly, the droplets were deformed being elongated at their rear end. Next, we perform experiments with  4.2 wt$\%$ PEO of M{$_{w,1}$}= 1 x 10$^6$ Da with surfactant concentration of 21 wt$\%$ (Fluid \rom{4}). In this solution with viscosity, $\eta_o$=13 Pa~s, the droplets propelled with speed $\sim$ 6-7 $\mu$m s$^{-1}$. Interestingly, in contrast to fluid \rom{3}, the droplets maintained their spherical shape during self-propulsion. Fig.\,\ref{trajectories} depicts the bright-field optical micrographs of the typical droplet trajectories in the different fluids, recorded for a time period of 100s, a duration for which the droplet size remains largely unchanged. An important observation here is that a noticeable droplet deformation was not observed for medium consisting of lower M$_w$ PEO, despite being viscosity-matched with that of high M$_w$ PEO. Also, for the medium with higher M$_w$ PEO, no deformation was observed at low droplet speeds. These observations strongly highlight the crucial role of droplet speed and polymer chain length dependent viscoelasticity in the observed deformation of the droplets.
 \vspace{0mm}

The Deborah number, $De$, quantifies the degree of the viscoelasticity in the surrounding medium. In order to quantify the deformation of active droplets, we define the deformation index ($\psi$) as the standard deviation of the distance of the points at the periphery of the droplet with respect to its geometrical center. Table \ref{systems} lists $U$, $\tau$, $De$ and the resulting $\psi$ values for the different fluids. Clearly, at low $De$, no deformation is observed and only beyond a moderate $De$, i.e., $De>0.20$, an increase in $\psi$ is observed.

   \begin{table} [ht]
   \caption{Characteristics of different ambient fluids.}
    \centering
    \begin{tabular}{ c  c  c  c  c}
    \hline
    Fluid & $ U $ & $ \tau $ & $ De $ & Deformation\\ [0.1ex]
   & $\mu$ms$^{-1}$ & s &  & Yes/No\\
    \hline 
    \rom{1}  & 15 & 0 & 0 & No\\
    \rom{2}  & 2 & 1 & 0.04 & No\\
    \rom{3}  & 10 & 1 & 0.20 & Yes\\
    \rom{4} & 6 & 0.025 & 0.003& No\\
   \rom{5}  & 2-16 & 2 &0.10-0.60 & Yes\\
    \hline
    \end{tabular}
    \label{systems}
    \end{table}
Although active droplets are known to be weak pushers, as a first approximation we assume the surrounding flow field and the resulting elastic stresses to be qualitatively given by a droplet driven by an external field. With this assumption, we determine the droplet shape following the methodology outlined in Noh \emph{et.al.}~\cite{doi:10.1063/1.858568}, where the deformation of a rising bubble was investigated in a polymeric solution. The viscoelastic fluid is approximated by the finitely extensible nonlinear elastic (FENE) dumbell model that was used by Chilcott and Rallison~\cite{CHILCOTT1988381}. The dumbells evolve according to
\begin{equation}
    \label{eq:FENE}
    \frac{\partial \mathbf{A}}{\partial t} + \mathbf{u}\cdot\nabla\mathbf{A} = \mathbf{A}\cdot\nabla\mathbf{u} + \nabla\mathbf{u}^{T}\cdot\mathbf{A} - \frac{f(\mathbf{R})}{De}(\mathbf{A}-\mathbf{I}) 
\end{equation}
where $\mathbf{A}$ is the ensemble average of the dyadic product $\mathbf{R}\mathbf{R}$ of the dumbell end-to-end vector $\mathbf{R}$. In this constitutive relation, $f(\mathbf{R}) = 1/(1-\mathrm{tr}(\mathbf{A})L^{-2})$ and $L$ is the maximum polymer chain length relative to the equilibrium radius of gyration. The polymer affects the shape of the drop through the normal stress balance at the interface, given by
\begin{equation}
    \label{eq:normalstress}
    -p+2E_{rr} +\frac{c_{PEO}}{De}f(\mathbf{R})A_{rr} = \frac{1}{Ca}\nabla\cdot\mathbf{n}
\end{equation}
for a slightly non-spherical drop. In Eq.\ref{eq:normalstress}, $E_{rr}$ is the radial component of the strain-rate tensor, $\mathbf{E} = (\nabla\mathbf{u}+\nabla\mathbf{u}^{T})/2$ and $Ca=\eta_{o} U/\gamma$ is the Capillary number, where $\eta_{o}$ is the bulk viscosity, $U$ is the droplet speed and $\gamma$ is the interfacial tension. The velocity profile is assumed to be the Stokes solution for flow around a spherical droplet. Thus, Eq. \ref{eq:FENE} yields a set of ordinary differential equations, which are solved numerically, to obtain the components of $\mathbf{A}$ as functions of $\theta$. 
\begin{figure}[H]
\centering
\includegraphics[width=0.9\linewidth]{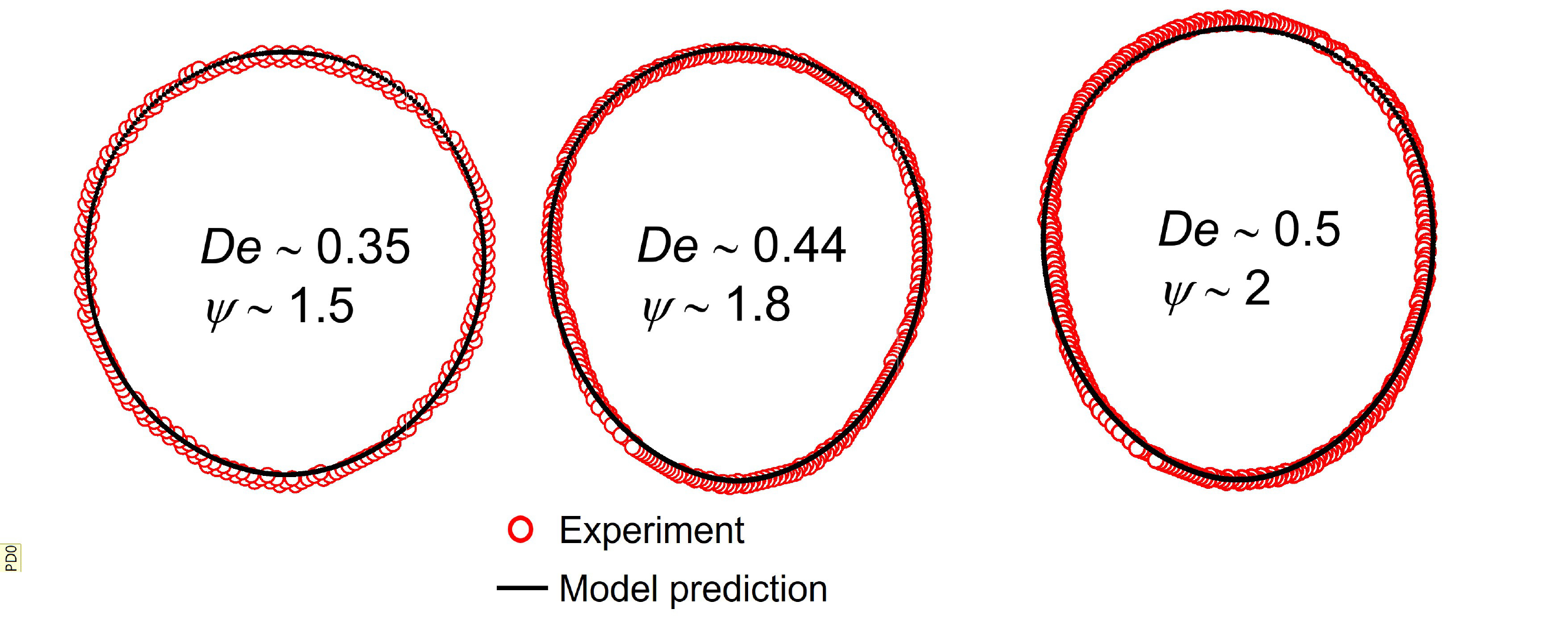}
\caption{Comparison of the experimentally obtained shape of the deformed droplets with respect to the model prediction at three different $De$. }
\label{fitting}
\end{figure}  

The approximate solution to a moving droplet with small deformation can be obtained from the normal stress balance as,
\begin{equation}
    \label{eq:radius}
    r(\theta) = 1 + c_{PEO}Ca\Sigma_{n=2}^{\infty}\beta_{n}P_{n}\cos(\theta),
\end{equation}
where $\beta_{n} = \frac{1}{(n-1)(n+2)(2n+1)}$\\$\times\int^{\pi}_{0}\frac{f(\mathbf{R})}{De}A_{rr}P_{n}(\cos\theta)\sin\theta d\theta$. Solving this equation, we obtain the radius as a function of $\theta$ for various values of $De$, $L$, $c$ and $Ca$.

The shape of the deformed droplets predicted by the model is compared with the experimentally observed shapes in fig.\,\ref{fitting}. A good match of the numerical prediction with that of the experimentally obtained shape confirms our hypothesis that the deformation of the droplet at the rear (close to rear stagnation point) arises due to the excess normal stress because of the stretching of polymer chains normal to the droplet surface.

\begin{figure}[H]
\centering
\includegraphics[width=0.6\linewidth]{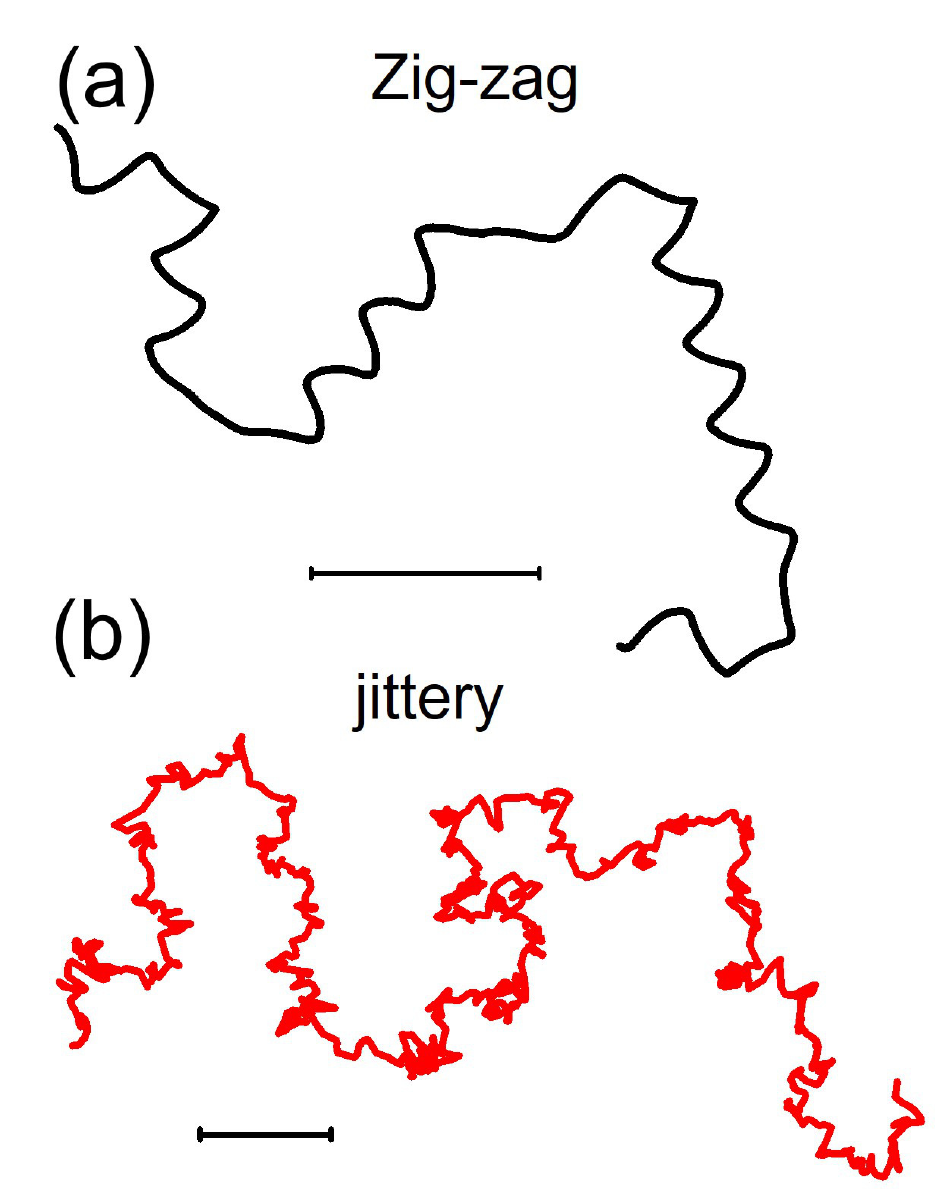}
\caption{(a) Representative trajectories of active 5CB droplets ($\sim$50 $\mu$m) performing (a) zigzag swimming motion in 21wt$\%$ aqueous TTAB solution doped with 1.25 wt$\%$ 8000 kDa PEO and (b) jittery motion in 80wt$\%$ glycerol aqueous solution containing 6 wt$\%$
TTAB. The scale bars correspond to 100 $\mu$m. }
\label{fig:traj}
\end{figure}

Further, to understand the effect of enhanced viscoelasticity, we repeated the experiments in fluid \rom{5}  (see Table \ref{systems}), where $c_{PEO}$ was increased to 1.25$\%$. In this medium, the active droplets demonstrated a periodic zigzag motion (fig. \ref{fig:traj}a). The droplets moved smoothly in a particular direction for some time followed by a change in direction of about $\sim$ 70-90$^{\circ}$. The droplet continues moving smoothly along the new direction for around the same time, until changing direction again, however this time in the opposite direction. It is to be noted that this zigzag motion of the droplet is different from the previously reported jittery motion \cite{doi:10.1063/5.0038716, PhysRevX.11.011043,Dwivedi2022arxiv} with repeated stop and go events (fig. \ref{fig:traj}b), wherein, the direction fluctuations are mostly at random angles. Additionally, in the jittery motion, the resting state is attributed to the non-propulsive quadrupolar flow-field mode, whereas, in the zigzag motion, no quadrupolar mode was observed (discussed later in the manuscript).

\begin{figure*}[htp]
\centering
\includegraphics[width=0.9\linewidth]{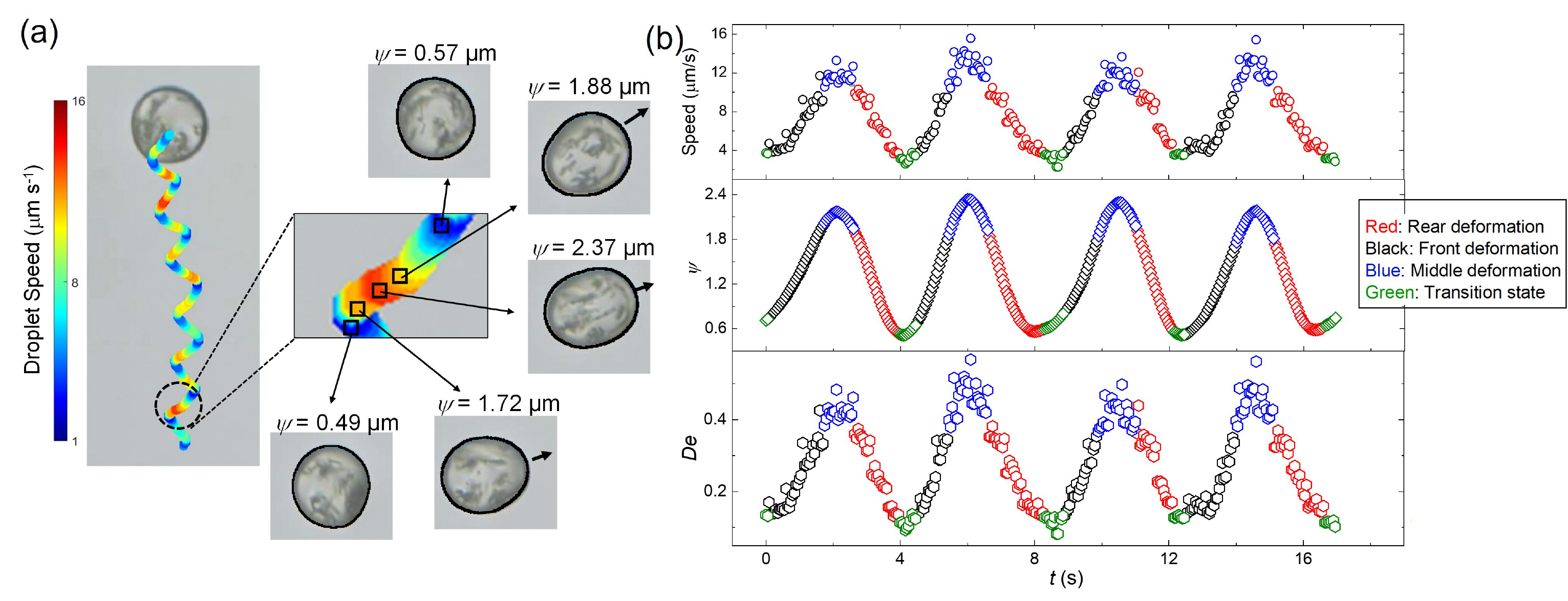}
\caption{(a) The X-Y trajectory, color code by propulsion speed, of an active 5CB droplet ($\sim$50$\mu$m) performing zigzag motion in 21wt$\%$ aqueous TTAB solution doped with 1.25 wt$\%$ 8000 kDa PEO. Enlarged insets highlight one section of trajectory between two consecutive turning events, with optical micrographs of droplets highlighting their deformation at different instances. Corresponding variation of (b) droplet speed (c) $\psi$ and (d) $De$ with time}
\label{cycles}
\end{figure*}

Fig. \ref{cycles}(a) depicts a representative section of the zigzag trajectory, color coded by the propulsion speed. The enlarged inset demonstrates the smooth intermediate section between two successive turning events of the periodic trajectory. The optical micrographs correspond to the droplet images captured at different positions. These images reveal a time-dependent shape deformation exhibited by the droplet. First, the droplet is at its lowest speed and is nearly spherical with minimum deformation. Subsequently, it accelerates while getting deformed at its front. Thereafter, with further increase in speed, at around half the time period, the deformation shifts to the equatorial region of the droplet rendering it elongated. Next, the droplet slows down, while getting deformed at its the rear end. Finally, the droplet regains its original shape with negligible deformation and slow speed, before switching the direction and repeating the cycle. Fig.\,\ref{cycles}(b) depicts the time-dependent speed ($U$), $\psi$ and $De$ for a few cycles. It is to be noted that the temporal variation in $U$, $\psi$ and $De$ are in phase with each other. The maximum deformation occurs at the highest speed leading to the highest $De$, wherein the droplet has an elongated shape. The least deformation is observed during the change in droplet direction, when the droplet is the slowest with the least $De$.

\begin{figure*}
\centering
\includegraphics[width=1\linewidth]{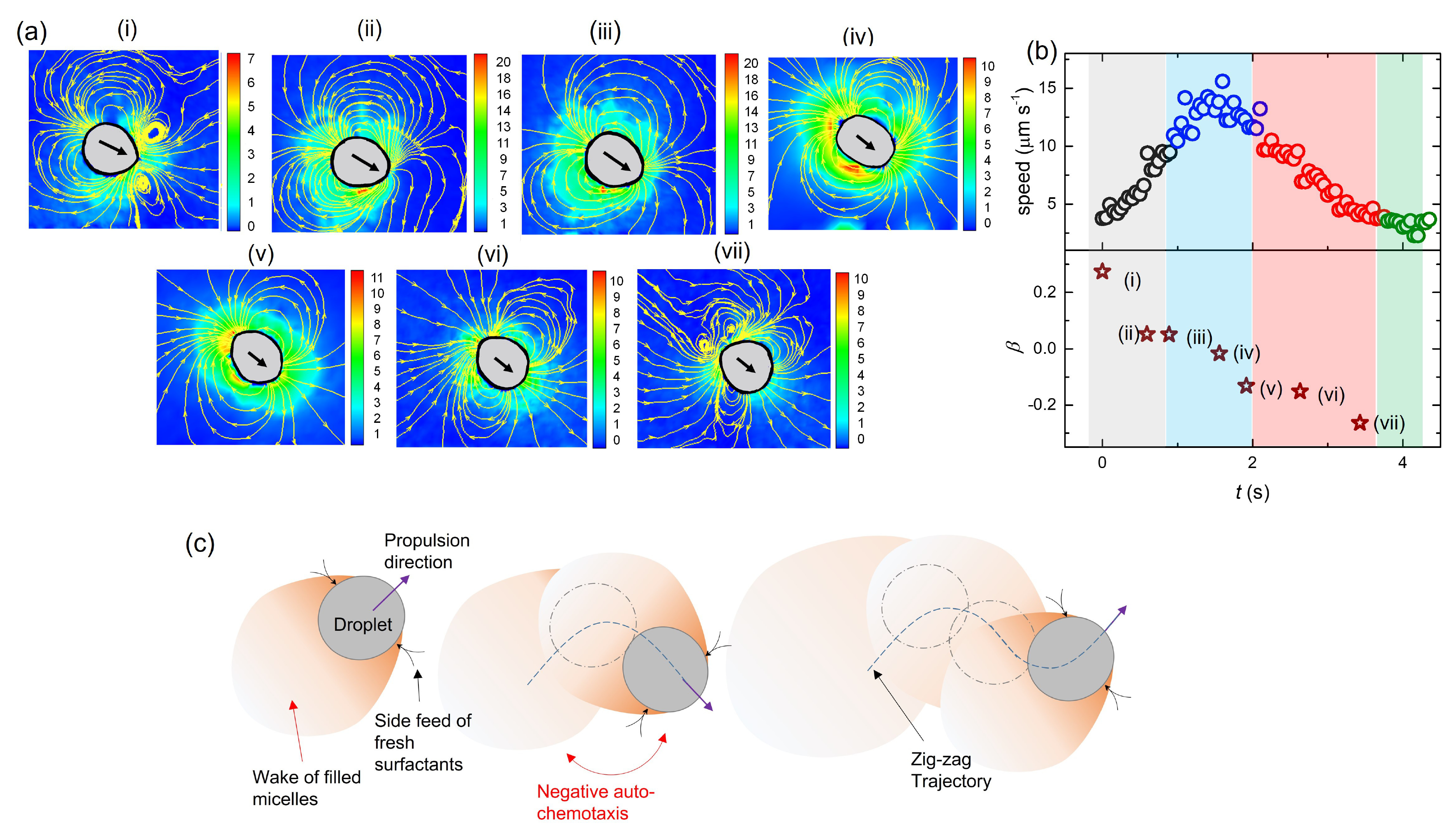}
\caption{(a) Fluorescent micrographs depicting streamlines obtained from PIV measurements. The streamlines represent the flow-field around the droplet in the laboratory frame for (i-iii) puller (iv) neutral and (v-vi) pusher swimming modes. Color bars represent the magnitude of the velocity field around the droplet. (b)  Variation of speed and $\beta$ with time during one persistent stretch in the zigzag cycle (c) Schematic depicting the mechanism of zigzag motion of the active 5CB droplet.}
\label{PIV}
\end{figure*}

By seeding the surrounding aqueous solution with fluorescent tracers (500 nm Polystyrene particles), we performed particle image velocimetry (PIV) to measure the fluid flow and identify the swimming gait of the droplet vis-à-vis neutral/pusher/puller ~\cite{https://doi.org/10.1002/cpa.3160050201}. In the laboratory frame, a pusher pulls the ambient fluid in from the equatorial region, and pushes them away from the front and the rear end. In contrast, in the puller mode, the fluid is pulled in from the front and the rear end, and pushed out from the equatorial region. The images shown in fig.\,\ref{PIV}(a(i-vii)) depicts the flow field around the droplet for one cycle of the zigzag motion. Initially, when the droplet is deformed from the front, the swimming mode is that of a weak puller. Thereafter, when the droplet is elongated from the equatorial region it transitions to a neutral swimmer. Eventually, while being deformed from the rear, it adopts the weak pusher mode, and finally comes to a near halt with a strong pusher mode. The PIV data was further used to compute the tangential velocity $u(R,\theta)$ at the droplet interface in the co-moving frame of reference. The droplet velocity at $r=R$ can be expressed as  
\begin{math}
        u(R,\theta)= \sum_{i=1}^{\infty}B_{n}V_{n}\cos(\theta) 
\end{math} 
, where
\begin{math}
        V_{n}\cos(\theta)=
        \frac{2}{n(n+1)}\sin(\theta)P_{n}\cos(\theta)
\end{math} 
and $P$$_{n}$ is the Legendre polynomials~\cite{blake1971spherical, lighthill1952squirming}. On assuming B$_{n}$=0 for $n>2$, $u(R,\theta)$ can be written as $\sim$
\begin{math}
        B_{1}\sin(\theta)+
        \frac{B_{2}}{2}\sin(2\theta).
\end{math}
The ratio, 
\begin{math}
        \beta=
        \frac{B_{2}}{B_{1}}
\end{math}
indicates the swimming mode. For $\beta >0$ it is puller, $\beta =0$ neutral and for $\beta <0$ it is the pusher mode ~\cite{pedley2016spherical}.

In fig.\,\ref{PIV}(b) we show the variation in propulsion speed and $\beta$ with time during one cyle of the zigzag motion. With time, $\beta$ decreases continuously showing a transition of swimming mode from puller to neutral to pusher. We also note that towards the end of the cycle, the pusher strength increases, which is characterized by the larger magnitude of $\beta$ ($<$0). It is to be noted that the onset of direction change happens only once the droplet adopts the pusher mode. The droplet motion remains persistent while it takes the other swimming modes at the beginning of the cycle. This behavior is in agreement with previous reports~\cite{Dwivedi2022arxiv, suda2021straight}, which demonstrated that the pusher mode is more susceptible to directional fluctuations. However, the question remains as to why the droplet follows a periodic direction change. Using the schematic shown in fig. 6(c), we try to explain this. In pusher mode, the droplet pulls in fresh fluid carrying a supply of fresh surfactant/empty micelles to the equatorial region. Therefore, following the chemotactic tendency the droplet turns to align its propulsion with the fresh feed. For the maiden cycle, the equatorial feed is expected to be axi-symmetric with respect to the propulsion direction, so the droplet can turn either way. During the subsequent cycles, due to the negative autochemotactic effect caused by the filled-micelles in its wake from previously occupied positions, the droplet steers in alternate directions, resulting in a periodic zigzag motion. Using numerical analysis, a similar zigzag motion was predicted by Li \cite{li_2022}. Depending on the Pecl\'et ($Pe$) and Damk\"ohler ($Da$) numbers in a Newtonian media, Li predicted that the interaction between the primary and secondary wakes results in the zigzag motion of the droplet. However, in our case the fundamental reason behind the observed oscillatory zigzag motion of droplet is the droplet deformation caused by the viscoelastic media.

In 2012, through numerical investigation of a squirmer in viscoelastic fluid, Zhu \emph{et.al.}~\cite{doi:10.1063/1.4718446} reported that the polymeric stress around a swimmer depends on the gait of swimming, i.e., pusher vs puller. The study demonstrated that the polymer chain conformation is strongly affected by the local flow field. In the pusher mode, the polymer chains are highly stretched at the rear end, generating large normal stresses. However, in the puller mode, the polymer chains are mainly stretched from the sides generating high normal stress zones in the equatorial region. The study assumes swimmers to be spherical at all times with an imposed mode of swimming, and therefore, no deformation in the swimmer shape was observed. Recently, using the minimum energy dissipation theorem ~\cite{PhysRevLett.126.034503} Daddi-Moussa-Ider \emph{et.al.} identified the optimum swimming mode depending on the shape of the swimmer ~\cite{daddi-moussa-ider_nasouri_vilfan_golestanian_2021}. The study predicted that for swimmmers with the deformation at the front, the puller mode is optimal. Whereas, for swimmers with deformation at the rear, the pusher mode is optimal. Our experimental observations are in line with these numerical predictions. For low $De$, the flow-field around the droplet ensures that polymer chains are highly stretched in the radial direction (larger values of $A_{rr}$ locally as captured by the model) close to the rear end. This results in additional normal stress at the droplet interface, causing the droplet to deform from the rear while adopting a pusher swimming mode. For higher $De$, in agreement with the prediction of Zhu \emph{et.al.}~\cite{doi:10.1063/1.4718446}, the local velocity field changes to stretch the polymer chains along the equatorial region and the swimming gait transitions to puller. The resulting extra normal stresses can produce noticeable local deformations at the front and the equatorial region. We suspect that the time dependent shape deformation originates due to the non-linear coupling of the hydrodynamic forces and the swimming dynamics. However, the exact reason is not yet clear and warrants more focused efforts in the future. 
    
In conclusion, we report the dynamics of deforming active droplets in viscoelastic media. The large scale deformation in the droplet shape is attributed to the extra normal stress localized at the droplet interface resulting from the extension of polymeric chains. We also demonstrate that the deformation is a function of the Deborah number, $De$, experienced by the droplet. We foresee that with more synchronous efforts from experiments, theory and simulations, a comprehensive understanding of the self-propulsion of active droplets in complex fluids can be achieved. While our observations are for synthetic active droplets, we envisage that it provides useful insights into the diverse range of active motion observed in nature including biological swimmers. 
 \begin{acknowledgments}
We acknowledge the funding received by the Science and Engineering Research Board (Grant numbers SB/S2/RJN105/2017 and ECR/2018/000401), the Department of Science and Technology, India. 
\end{acknowledgments}

\bibliography{apssamp}

\end{document}